\documentclass[prl,twocolumn,showpacs,amsmath]{revtex4}
\usepackage{graphicx}%
\usepackage{dcolumn}
\usepackage{amsmath}
\usepackage{here}

\newcommand\pictc[5]{\begin{figure}
                       \centerline{
                       \includegraphics[width=#1\textwidth]{./#3}}
                   \vspace*{-0.1cm}\protect\caption{\protect\label{fig:#4} #5}
                    \vspace*{-0.05cm}\end{figure}            }
\newcommand\pict[4][0.45]{\pictc{#1}{!tb}{#2}{#3}{#4}}

\newcounter{Fig}

\begin{document}

\title{Observation of discrete vortex solitons in optically-induced photonic lattices}

\author{Dragomir~N. Neshev}
\affiliation{Nonlinear Physics Group, Research School of Physical Sciences and Engineering, Australian National University, Canberra ACT 0200, Australia}

\author{Tristram~J. Alexander}
\affiliation{Nonlinear Physics Group, Research School of Physical Sciences and Engineering, Australian National University, Canberra ACT 0200, Australia}

\author{Elena~A. Ostrovskaya}
\affiliation{Nonlinear Physics Group, Research School of Physical Sciences and Engineering, Australian National University, Canberra ACT 0200, Australia}

\author{Yuri~S. Kivshar}
\affiliation{Nonlinear Physics Group, Research School of Physical Sciences and Engineering, Australian National University, Canberra ACT 0200, Australia}

\author{Hector Martin}
\affiliation{Department of Physics and Astronomy, San Francisco State University, CA 94132}
\affiliation{TEDA College, Nankai University, Tianjin, China}

\author{Igor Makasyuk}
\affiliation{Department of Physics and Astronomy, San Francisco State University, CA 94132}
\affiliation{TEDA College, Nankai University, Tianjin, China}

\author{Zhigang Chen}
\affiliation{Department of Physics and Astronomy, San Francisco State University, CA 94132}
\affiliation{TEDA College, Nankai University, Tianjin, China}

\begin{abstract}
We report on the first experimental observation of {\em discrete vortex solitons} in two-dimensional optically-induced photonic lattices. We demonstrate strong stabilization of an optical vortex by the lattice in a self-focusing nonlinear medium and study the generation of the discrete vortices from a broad class of singular beams.
\end{abstract}

\pacs{  42.65.Tg,   %Optical solitons; nonlinear guided waves
        42.65.Jx,   %Beam trapping, self-focusing and defocusing; self-phase modulation
        42.70.Qs    %Photonic bandgap materials
}

\maketitle

Periodic photonic structures and photonic crystals recently attracted a lot of interest due to the unique ways they offer for controlling light propagation. Periodic modulation of the refractive index modifies the diffraction properties and strongly affects nonlinear propagation and localization of light~\cite{book}. Recently, many nonlinear effects including the formation of lattice solitons have been demonstrated experimentally in one- and two-dimensional optically-induced photonic lattices~\cite{FleisPRL,NeshOL,FleisNature,Martin}. The concept of optically-induced lattices ~\cite{efrem} relies on the modulation of the refractive index of a nonlinear medium with periodic optical patterns, and the use of a weaker probe beam to study scattering of light from the resulting periodic photonic structure.

So far, only simple stationary structures have been described theoretically and generated experimentally in optically-induced lattices~\cite{FleisPRL,NeshOL,FleisNature,Martin,efrem}. One of the most important next steps is the study of nonlinear modes with a nontrivial phase such as {\em vortices}, the fundamental localized objects appearing in many branches of physics. In optics, vortices are associated with the screw phase dislocations carried by diffracting optical beams~\cite{soskin}. When such vortices propagate in a {\em defocusing} nonlinear Kerr-like medium, the vortex core becomes self-trapped, and the resulting structure is known as {\em an optical vortex soliton}~\cite{book}. Such vortex solitons are usually generated experimentally on a broad background beam~\cite{swartz,chen97}. They demonstrate many similarities with the vortices observed in superfluids and Bose-Einstein condensates.

In contrast, optical vortex solitons {\em do not exist in a self-focusing} nonlinear medium; a ring-like optical beam with a phase dislocation carrying a finite orbital angular momentum~\cite{kruglov} decays into the fundamental solitons flying off the main ring~\cite{Firth:1997-2450:PRL}. This effect was first observed experimentally in saturable Kerr-like nonlinear medium~\cite{tikh}, and then in photorefractive~\cite{chen97} and quadratic~\cite{chi2} nonlinear media in the self-focusing regime.

Recent theoretical studies of the discrete~\cite{vort_discrete} and continuous models of nonlinear periodic lattices~\cite{ziad,Baizakov:2003-642:EPL} suggest that the vortex-like structures can be supported by the lattice even in the self-focusing regime. In this Letter, we report on the first experimental observation of {\em discrete (lattice) vortex solitons} and demonstrate, both theoretically and experimentally, that localized optical vortices can be generated in a self-focusing nonlinear medium, being stabilized by the two-dimensional periodic potential of a photonic lattice.

To lay a background for our experiment, first we study numerically the generation of discrete vortex solitons in a two-dimensional photonic lattice optically-induced in a photorefractive crystal. Since in a typical experiment the input beam is radially-symmetric and does not posses the shape and symmetry of a stationary lattice vortex~\cite{ziad,Baizakov:2003-642:EPL}, it is crucially important to understand whether the generation of such a state is possible from a broad range of initial conditions.

A two-dimensional square lattice created by ordinary polarized beams can be described by the intensity pattern $I_g(x,y) = I_0 \,{\rm sin}^2(\pi x/d){\rm sin}^2(\pi y/d)$, where $I_0$ is the maximum lattice intensity, and $d$ is the period [Fig.~\ref{fig:fig1}(d)]. Within the approximation of isotropic photorefractive nonlinearity, the evolution of a extraordinary probe beam $u$ is governed by the nonlinear equation~\cite{efrem}
\begin{equation}
\label{norm}
i\frac{\partial u}{\partial z} + \frac{1}{2} \Delta_{\perp} u
    - \frac{V_0 u}{1 + I_g(x,y) + |u|^2} = 0,
\end{equation}
where $\Delta_{\perp}$ is the two-dimensional Laplacian, the intensities are normalized to the dark irradiance $I_d$ of the crystal, $V_0=\gamma_{nl}x^2_0/2$ ($V_0 = 1$ corresponds to a bias field of $74.6$~V/cm), $\gamma_{nl} \equiv k_0^2 n_e^4 r_{33}{\cal E}_0$, and $n_e$ is the refractive index for extraordinary polarized beams. For our choice of polarity of the bias field, ${\cal E}_0$, nonlinearity exhibited by the probe beam is {\em self-focusing}. The transverse coordinates are measured in the units of $x_0=d/2$, and the propagation distance, in units of $k_0n_ex_0^2$. In our simulations we assume normalization corresponding to the experimental parameters. Consequently, $x = 1$ corresponds to 14~$\mu$m and $z = 1$ corresponds to propagation distance of 5.86~mm.

To answer the question of stable discrete-vortex generation from a radially-symmetric input beam carrying {\em a screw phase dislocation} of unit topological charge, we study numerically a wide range of initial beam widths and intensities. At narrow widths, the spectrum of the initial beam is wide and may couple strongly to the higher-order spectral bands, leading to rapid diffraction and break-up of the initial beam.  Broad beams will instead populate many lattice sites, and so result in a beam structure far from any stable stationary configurations.
%However at beam widths of the order of the lattice period, the nonlinearity may detune the spectrum into the semi-infinite band gap, away from the linear transmission bands and so allow stable generation of solitons.
Generation therefore requires a balance between the nonlinearity and beam diffraction properties for coupling to one of the stationary vortex states on the lattice.

Several different types of discrete vortices can be supported by the lattice. The lowest-order symmetric {\em off-site} [Fig.~\ref{fig:fig1}(f)] and {\em on-site} (not shown) vortex states were first suggested theoretically in Refs.~\cite{ziad,Baizakov:2003-642:EPL} for a vortex centered between four sites of the lattice or on a single lattice site, respectively. Asymmetric stationary vortex states can also be supported by the lattice. The off-site discrete vortex shown in Fig.~\ref{fig:fig1}(f), however, represents the state with {\em the strongest coupling} between its lobes. It can be generated from a broader class of initial conditions, which facilitates its experimental observation.
%We have optimized the initial conditions for the generation of this off-site vortex for beams centered 'off' the lattice, i.e. on a lattice minimum.
For the given experimental conditions of a saturation corresponding to $I_0 = 10$ and an initial beam peak intensity of roughly $I_0/5$, we find the optimal generation conditions for an off-site vortex are given by an input beam of the form $u(r,\theta) = a\, r \exp(-w r^2 + i\theta)$, where $(r,\theta)$ are the polar coordinates, $a = 3.3$, and $w = 1$ [Fig.~\ref{fig:fig1}(a)].

%--------------------------------------------------------------------------
\pict{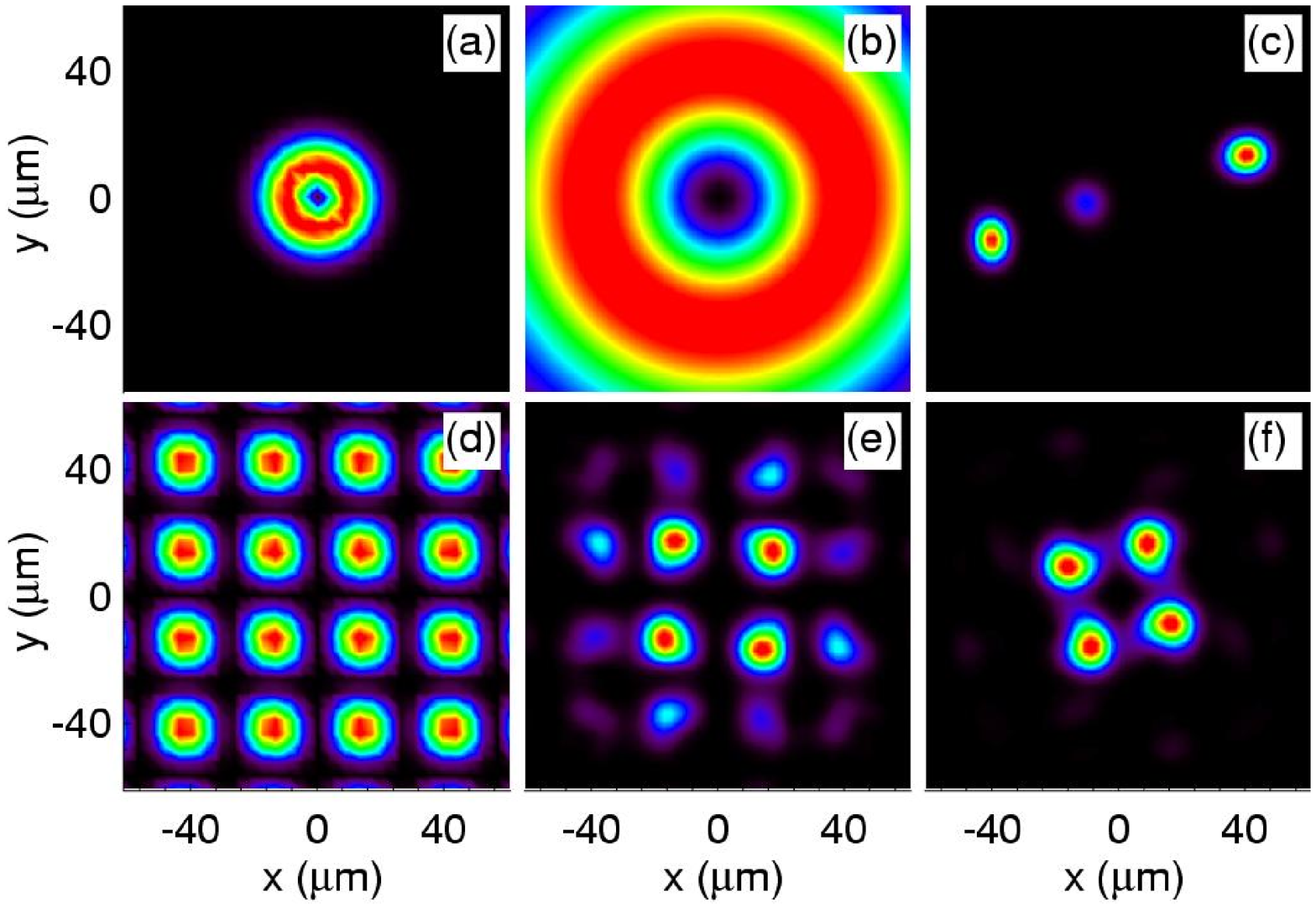}{fig1}{Numerical results. (a) Input vortex and (d) linear lattice, $I_0 = 10 I_d$. (b) Linear diffraction at $z=2$ and (c) nonlinear propagation at $z=10$ in a homogeneous medium. (e,f) discrete diffraction and discrete vortex soliton on the lattice, respectively at $z=10$ and $V_0 = 16.09$.}
%--------------------------------------------------------------------------

First, we simulate the vortex propagation without the lattice. In the absence of nonlinearity, the input beam diffracts, as shown in Fig.~\ref{fig:fig1}(b) for $z=2$. In the nonlinear regime, the vortex decays into a diverging pair of out-of-phase solitons [Fig.~\ref{fig:fig1}(c)]. This type of azimuthal instability is generic for vortex-carrying beams, however the exact number of filaments depends on the input beam parameters~\cite{Firth:1997-2450:PRL}. The filaments diverge due to the angular momentum carried by the initial vortex beam~\cite{chen97,tikh,chi2}.

In the presence of a two-dimensional lattice and at low probe-beam intensities (or $|u|^2$ in Eq.~\ref{norm} neglected), the input beam exhibits {\em discrete diffraction}, whereby its power tends to re-distribute between the neighboring lattice sites, as shown in Fig.~\ref{fig:fig1}(e) for relatively long propagation distance $z=10$. Remarkably, in the nonlinear regime, the input beam does not decay or diffract due to decoupling from the linear transmission spectrum, and instead transforms into a four-lobe structure [Fig.~\ref{fig:fig1}(f)]. It corresponds to a {\em discrete vortex soliton}~\cite{vort_discrete,ziad}, which propagates stably in the lattice even under strong nonlinear perturbations induced by the transient `breathing' mode, seen as slight rotational oscillations in the numerical simulations [Fig.~\ref{fig:fig1}(f)]. This stable vortex state is generated by centering the initial vortex beam between four sites of the optical lattice, and therefore corresponds to an {\em off-site} discrete vortex. 

%--------------------------------------------------------------------------
\pict[0.48]{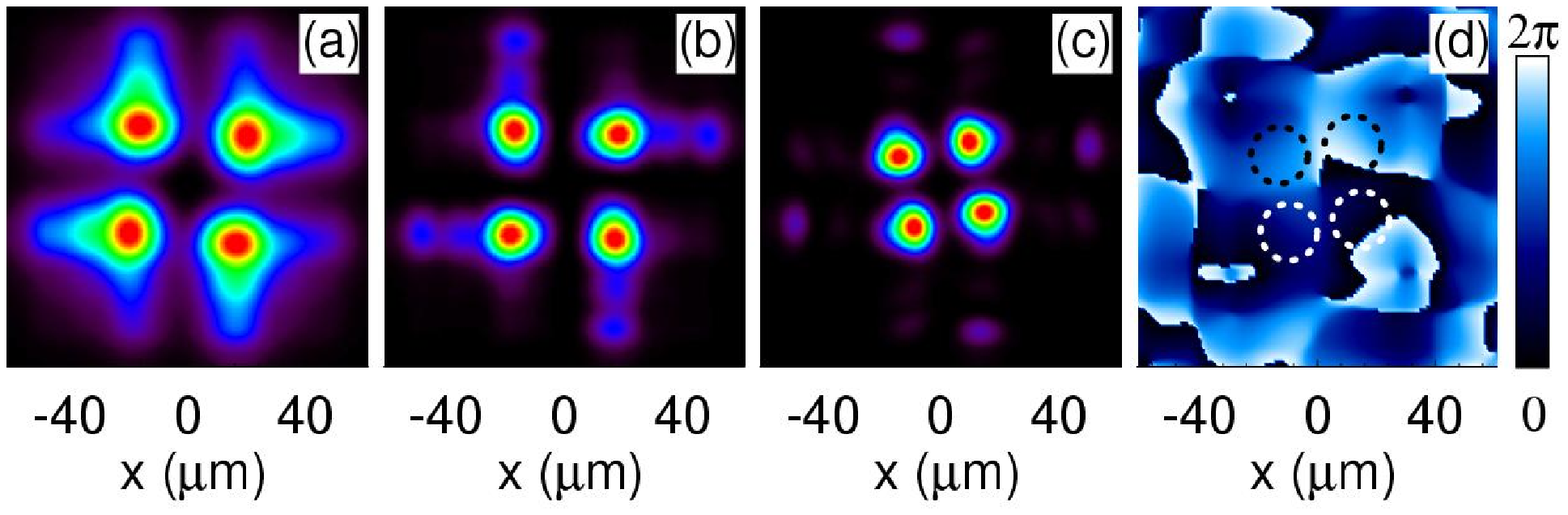}{fig2}{Dependence of the vortex state at the crystal output ($z=8.2$~mm, $I_0 = 10I_d$) on the strength of the nonlinearity, $V_0$ (or bias field). (a-c) Intensity distribution for $V_0 = 2.7$, $V_0 = 8.04$ and $V_0 = 16.09$, respectively. (d) phase distribution corresponding to the stationary state (c). Dashed lines correspond to the position of the four intensity lobes. Side-bar - phase colour map.}
%--------------------------------------------------------------------------

Next, we study the localization of the off-site discrete vortex for different strengths of nonlinearity (bias field) [Figs.~\ref{fig:fig2}(a-c)]. At low bias field, the focusing effect of nonlinearity is not sufficient to form a discrete vortex soliton, and the beam diffracts on the lattice [(a)]. A discrete vortex state is generated for stronger bias fields [(b,c)], and its confinement by the lattice varies with nonlinearity.
%Here, the four lobes of the discrete vortex are trapped by the closest neighboring minima of the potential, and the structure is stabilized by the lattice.
In order to verify that the structure generated by the input singular beam on the lattice is a discrete vortex, and not four uncoupled solitons, we analyzed the phase structure of the output state. In full agreement with earlier analysis~\cite{vort_discrete,ziad}, the generated discrete vortex soliton is composed of four phase-correlated lobes with a total phase ramp of $2\pi$ [Figs.~\ref{fig:fig2}(d)]. The observed phase structure closely resembles that of a soliton cluster~\cite{Desyatnikov:2002-53901:PRL}, with the main difference that the discrete vortices are robust and stationary structures.

%%%%%%  experimental results  %%%%%%

%--------------------------------------------------------------------------
\pict[0.45]{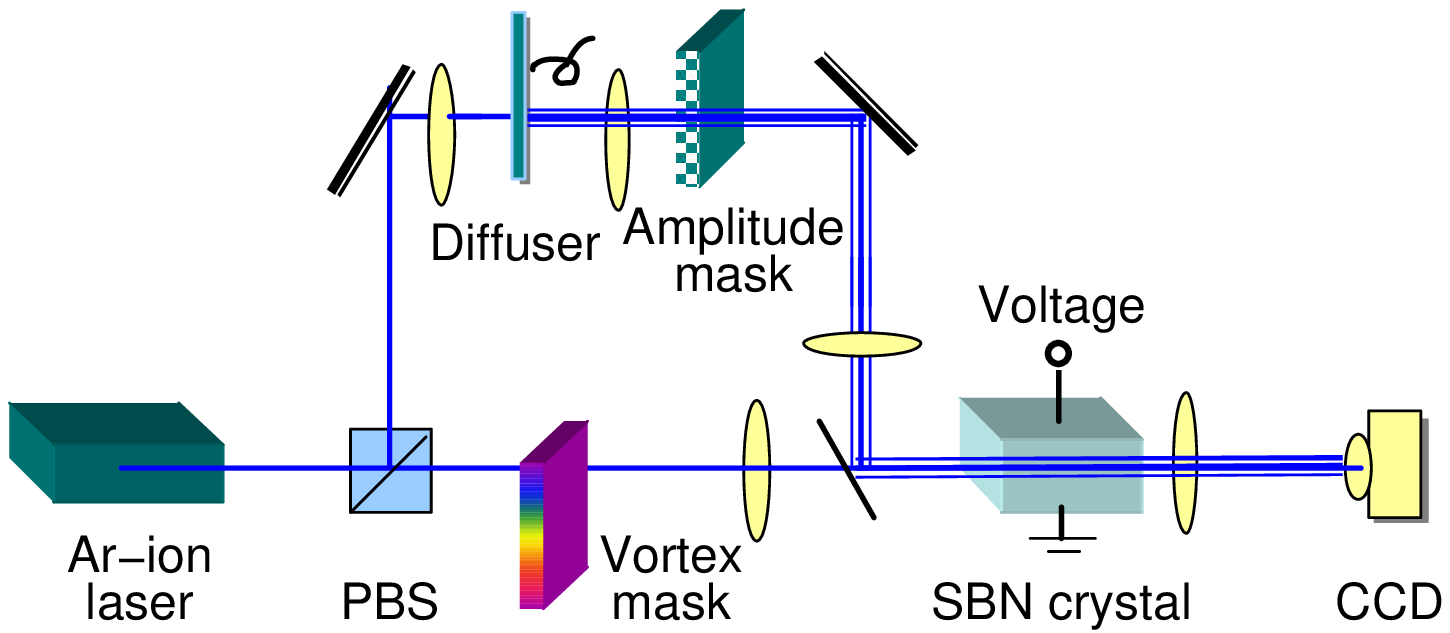}{setup}{Experimental setup. PBS: Polarizing beam splitter, SBN: Strontium Barium Niobate crystal.}
%--------------------------------------------------------------------------

Finally, we demonstrate the generation of the discrete vortices in experiment. Our experimental setup (Fig.~\ref{fig:setup}) is similar to that reported earlier in Refs.~\cite{chen02,Martin}. An argon ion laser beam (at the wavelength $\lambda=488$ nm) is collimated and then split with a polarizing beam splitter. The ordinarily-polarized beam (o-beam) is focused onto a rotating diffuser, turning into a partially spatially incoherent source. A biased photorefractive crystal (SBN:$60$, $5 \times  5 \times 8$~mm$^3$, with $r_{33}=280$~pm/V) is employed to provide a self-focusing non-instantaneous nonlinearity as in experiments with partially coherent solitons~\cite{incoh}.

To generate a two-dimensional photonic lattice, we use {\em an amplitude mask} and modulate spatially the otherwise uniform o-beam after the diffuser. The mask is then imaged onto the input face of the crystal, thus creating a partially coherent pixel-like input intensity pattern~\cite{chen02}. The extraordinarily-polarized probe beam (e-beam) is sent through a transmission vortex mask with unit topological charge, and then focused onto the crystal input face, propagating collinear with the lattice. In addition, a uniform incoherent background o-beam (not shown) is used as ``dark illumination" for fine-tuning the nonlinearity~\cite{incoh}. The input and output faces of the crystal are imaged onto a CCD camera and the beam separation was achieved by blocking one of the components and quickly recording the other one.

%--------------------------------------------------------------------------
\pict[0.48]{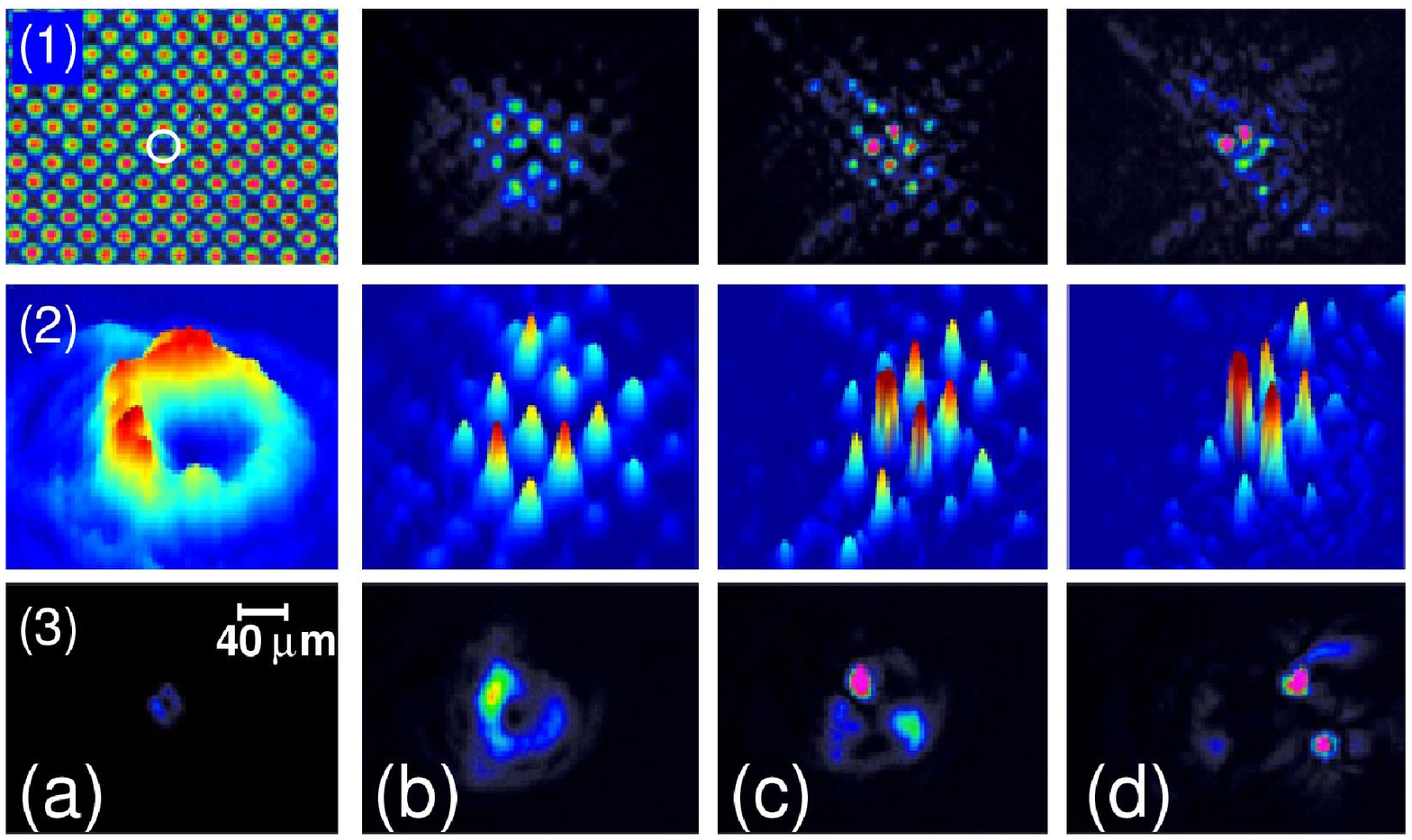}{experiment}{Experimental observation of an optical vortex propagating with (rows 1,2) and without (row 3) an optically-induced lattice. (a-1) the input pattern of the lattice, where the ring indicates the input vortex location; (a-2) and (a-3) the vortex beam at crystal output and input without bias field; (b-d) the vortex beam at output for bias fields of 600, 1200, and 3000~V/cm, respectively. Row~2 -- three dimensional plots of vortex beam intensity.}
%--------------------------------------------------------------------------

Our typical experimental results are summarized in Fig.~\ref{fig:experiment}. A two-dimensional square lattice is created first, with the principal axes oriented in the diagonal directions, as shown in Fig.~\ref{fig:experiment}(a-1). Since the lattice has a spatial period of only 28~$\mu$m, this orientation favors its stable formation~\cite{Martin}. 
%(Orientation in the $x-y$ directions tends to cause distortion due to anisotropic photorefractive nonlinearity.) 
In addition, the lattice beam is {\em partially spatially incoherent} (spatial coherence length $\sim$100~$\mu$m), which also entails stable lattice formation due to suppression of incoherent modulation instability~\cite{MarinPRL00}. The resulting periodic intensity pattern acts as an photonic lattice, with the lattice sites (intensity maxima) corresponding to the minima of the periodic potential experienced by the probe beam.

The input vortex beam, shown in Fig.~\ref{fig:experiment}(a-3), is then launched straight between four lattice sites, as indicated by a bright ring in the lattice pattern (a-1). The vortex beam has an intensity about 5 times weaker than that of the lattice. In this optical medium, the o-polarized lattice exhibits only a weak nonlinearity as compared to that experienced by the e-polarized vortex beam~\cite{efrem} and, therefore, it remains nearly invariant as the bias field increases, with only a slight increase of its intensity contrast. In agreement with our numerical modeling, the vortex beam exhibits discrete diffraction when the nonlinearity is low, whereas it forms a discrete vortex soliton at a higher nonlinearity. In Fig.~\ref{fig:experiment}(b-1,2), discrete diffraction of the vortex beam is shown at a low bias field of 600~V/cm. The symmetry of the observed diffraction pattern is distorted due to slight asymmetry of the input vortex profile and inherent defects inside the crystal. When the bias field is increased to 1200~V/cm [Fig.~\ref{fig:experiment}(c-1,2)], partial focusing of the vortex beam is observed. In this case, more energy of the vortex beam goes to the central four cites, but the side lobes still share a significant amount of the energy. Importantly, for a higher nonlinearity, i.e. at a bias field of 3000~V/cm, a discrete vortex soliton is clearly observed [Fig.~\ref{fig:experiment}(d-1,2)], with most of the energy concentrated at the central four sites along the principal axes of the lattice, indicating that a balance has been reached between the discrete diffraction and self-focusing experienced by the vortex.

As predicted by the theory, the observed discrete diffraction and self-trapping of the vortex beam in the photonic lattice is remarkably different from that in a homogeneous medium. In Fig.~\ref{fig:experiment} (row~3), we show the output intensity patterns of the vortex beam without the lattice, but for the same bias field (nonlinearity). At a low bias of 600~V/cm, the vortex beam is slightly modulated, but it maintains a doughnut-like diffraction pattern [Fig.~\ref{fig:experiment}(b)]. At a higher bias of 1200~V/cm, the vortex beam breaks up into several filaments due to the azimuthal instability in the self-focusing crystal [Fig.~\ref{fig:experiment}(c)]. After the break-up, the filaments tend to form solitons and move away from their original locations [Fig.~\ref{fig:experiment}(d)] towards the direction of crystalline $c$-axis due to the diffusion-induced self-bending enhanced by the high bias field~\cite{singh}. In contrast, the vortex beam is well trapped by the optical lattice, which suppresses the rich instability-induced dynamics (rotating, diverging, and self-bending) of the filaments in a homogeneous nonlinear medium.

%--------------------------------------------------------------------------
\pict[0.49]{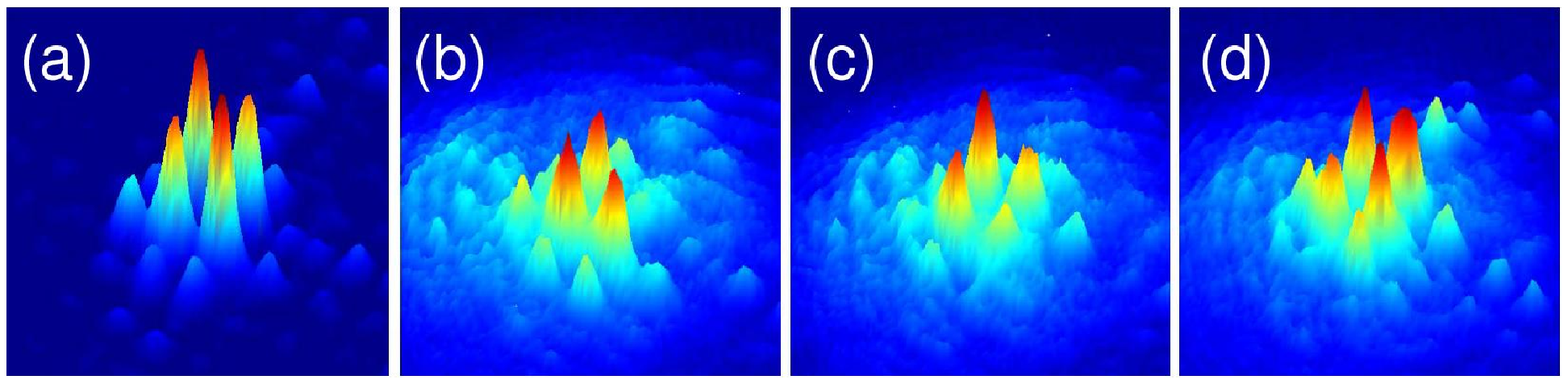}{interferograms}{Phase structure measurements. (a) Discrete vortex soliton for a lattice period of 20~$\mu$m. (b-d) Interferograms of the vortex soliton with a weak broad coherent beam, whose relative phase is changed in steps of $\pi/2$.}
%--------------------------------------------------------------------------

The observed discrete vortex soliton is a robust structure, which can be reproduced with different experimental parameters such as lattice spacing. In Fig.~\ref{fig:interferograms}(a) we show another example of the discrete vortex soliton formed on a lattice with a period of 20~$\mu$m. The four-lobe output intensity pattern was stable against ambient perturbations such as air vibration. To verify the nontrivial phase structure of the discrete vortex, we set up a Mach-Zehnder interferometer with a piezo-transducer (PZT) mirror in the reference-beam path. A weak quasi-plane wave beam was introduced for interference with the vortex after it exited the crystal. We then actively varied the relative phase between the plane wave and the vortex beam by the PZT mirror, and obtained a series of interferograms for testing of the phase distribution of the vortex. In this way, we avoid errors coming from the inherent noise in the reference beam phase front. Three of such interferograms taken as the mirror was driven gradually towards one direction are presented in Fig.~\ref{fig:interferograms}(b-d), which show clearly (see colour plots) that one of the four lobes increases its intensity whereas the intensity of the corresponding diagonal lobe decreases. Furthermore, the lobe with the strongest intensity is alternating among the four lobes. From these interferograms, we confirmed that the four lobes resulting from a discrete vortex soliton maintain a nontrivial phase relation. Since a vortex with a unit topological charge has a total phase ramp of $2\pi$, the relative phase between the four lobes changes in steps of $\pi/2$.

%The observed discrete vortex soliton is a robust structure, which can be generated for a variety of experimental parameters, including different lattice spacing. In Fig.~\ref{fig:interferograms}(a) we show another example of the discrete vortex soliton formed on a lattice with a period of 20~$\mu$m. The four-lobe output intensity pattern has been tested to be stable against ambient perturbations.

%To verify the nontrivial phase of the observed four-lobe state, we introduced a weak coherent broad reference beam and interfered it with the discrete vortex after it exits the crystal. We used a piezo-transducer mirror in the reference-beam path in order to control its phase relative to the vortex beam and obtained four interference patterns in phase steps of $\sim\pi/2$. Such {\em four-frame interferogram technique} allows for full reconstruction of the phase distribution of the vortex. In this way, we avoid any inherent curvature or noise of the reference beam phase front. Three of the interferograms are presented in Fig.~\ref{fig:interferograms}(b-d), which show that one of the four lobes increases its intensity whereas the intensity of the corresponding diagonal lobe decreases. Furthermore, the lobe with the strongest intensity is alternating among the four spots. From these interferograms, we have rigorously confirmed that the output state has a nontrivial phase relation and a total {\em phase ramp}, as expected for the discrete vortex soliton.

To confirm that the vortex trapping in the lattice is a result of the nonlinear self-action rather than simply due to the increased depth of the lattice potential, we perform additional experiments, where we significantly lower the intensity of the input beam while keeping the bias field unchanged. In this case we observe that the vortex experiences discrete diffraction similar to Fig.~\ref{fig:experiment}(b-1,2), as more energy of the vortex beam transfers further away from its center.

In conclusion, we have studied numerically the propagation of an optical beam with a phase dislocation in a two-dimensional periodic photonic structure, and predicted the stabilizing effect of the lattice on the vortex. We have observed experimentally the formation of a robust {\em discrete optical vortex} in two-dimensional optically-induced photonic lattices. The vortex has been generated in a self-focusing nonlinear medium for a variety of input conditions and lattice parameters. The experiments confirmed that the lattice stabilizes the vortex beam in the form of a non-diffracting four-lobe stationary structure with a screw phase dislocation. Our results can be useful for other branches of physics such us the nonlinear dynamics of Bose-Einstein condensates in optical lattices, where similar structures can be predicted theoretically and observed in experiment.

This work was supported by the Australian Research Council, U.S. AFOSR, and Research Corp. We thank A.~A. Sukhorukov for valuable discussions.

\end{document}